\documentclass[superscriptaddress,aps,pra,10pt,twocolumn,showpacs,notitlepage,nofootinbib,longbibliography]{revtex4-2}
\usepackage[utf8x]{inputenc}
\usepackage{amsmath,amsthm}
\usepackage{dcolumn}
\usepackage{bm}
\usepackage[colorlinks=true,citecolor=blue,urlcolor=blue]{hyperref}
\usepackage{graphicx}
\usepackage{subcaption}
\usepackage{braket}
\usepackage{color}
\usepackage[T1]{fontenc}
\usepackage[usenames,dvipsnames,svgnames,table]{xcolor}
\usepackage{lipsum}
\allowdisplaybreaks

\newcommand{\be}{\begin{equation}}
\newcommand{\ee}{\end{equation}}
\newcommand{\ba}{\begin{eqnarray}}
\newcommand{\ea}{\end{eqnarray}}

\newtheorem{cor}{Corollary}

\newtheorem{thm}{Theorem}
\newtheorem{Lemma}{Lemma}

\begin{document}

\title{Ability of unbounded pairs of observers to achieve quantum advantage in random access codes with a single pair of qubits}

\author{Debarshi Das}
\email{dasdebarshi90@gmail.com}
\affiliation{S. N. Bose National Centre for Basic Sciences, Block JD, Sector III, Salt Lake, Kolkata 700 106, India}

\author{Arkaprabha Ghosal}
\email{a.ghosal1993@gmail.com}
\affiliation{Centre for Astroparticle Physics and Space Science (CAPSS), 
Bose Institute, Block EN, Sector V, Salt Lake, Kolkata 700 091, India}

\author{Ananda G. Maity}
\email{anandamaity289@gmail.com}
\affiliation{S. N. Bose National Centre for Basic Sciences, Block JD, Sector III, Salt Lake, Kolkata 700 106, India}

\author{Som Kanjilal}
\email{som.kanjilal1011991@gmail.com}
\affiliation{Harish-Chandra Research Institute, Chhatnag Road, Jhunsi, Prayagraj (Allahabad) 211 019, India}

\author{Arup Roy}
\email{arup145.roy@gmail.com}
\affiliation{Department of Physics, A B N Seal College, Cooch Behar, West Bengal 736 101, India}

	\begin{abstract}
	Complications in preparing and preserving quantum correlations stimulate recycling of a single quantum resource in information processing and communication tasks multiple times. Here, we consider a scenario involving multiple independent pairs of observers acting with unbiased inputs on a single pair of spatially separated qubits sequentially. In this scenario, we address whether more than one pair of observers can demonstrate quantum advantage in some specific $2 \rightarrow 1$ and $3 \rightarrow 1$ random access codes. Interestingly, we not only address these in the affirmative, but also illustrate that unbounded pairs can exhibit quantum advantage. Furthermore, these results remain valid even when  all observers perform suitable projective measurements and an appropriate separable  state is initially shared.
	\end{abstract}
	\maketitle
	
	\section{Introduction}\label{s1}

Random access code (RAC) \cite{raco1,raco2,raco3} is one
of the fundamental communication protocols which, when assisted with quantum resources, manifests the astonishing potential of quantum systems in the context of information processing. In a $n \rightarrow m$ RAC, $n$ is the number of bits ($x_0$, $x_1$, $\cdots$, $x_{n-1}$)  accessed by the sender, say, Alice. On the other hand, $m$ is the number of bits that Alice is allowed to send the receiver, say, Bob with $m < n$.  In each run, Bob chooses the number $y$ randomly (where $y \in \{0, 1, \cdots, n-1\}$) and tries to guess the bit $x_y$ accessed by Alice, but unknown to Bob. The efficacy of RAC is limited when only classical strategies are employed. However, one can surpass the best classical strategies using quantum resources, e.g., by  using either quantum communication \cite{raco3} or classical bit communications  assisted with a shared bipartite quantum state \cite{erac,srac1}. 

RAC assisted with quantum resources was initially introduced \cite{raco1,raco2,raco3} in order to demonstrate the immense capabilities of quantum  systems in information processing tasks. The state of a $m$-qubit system can be represented by a unit vector in a $2^m$ dimensional complex Hilbert space, which opens up the possibility of encoding and transmitting classical information with exponentially fewer qubits, for example, Alice encoding $n$ bits into a $m$-qubit system (where $n>>m$) and sending it to Bob. However, due to the Holevo bound \cite{holevo}, $m$ qubits cannot transmit more than $m$ classical bits of information faithfully. Hence, it can be inferred that exponentially many degrees of freedom of a quantum system remain inaccessible. Nevertheless, the situation becomes interesting when Bob does not need to know all the $n$ bits of information together and chooses which bit of classical information he would like to extract out of the encoding. In order to extract different bits of information, Bob performs different measurements and  these measurements are in general not commuting. Thus, by choosing a particular measurement, Bob inevitably disturbs  the state and destroys some or all the information that would have been revealed by other possible measurements. This leads to the idea of RAC assisted with quantum resources. RAC has served as a powerful quantum communication task with various applications ranging from quantum finite automata \cite{raco2,raco3,Nayak99}, communication complexity \cite{ccx,Aaronson04,Gavinsky06,Buhrman01,Mar18}, non-local games \cite{Bridge}, network coding \cite{netcod1,netcod}, locally decodable codes \cite{ldc1,ldc2,ldc3}, dimension witnessing \cite{dw,dw2,dw3,dw4}, quantum state learning \cite{Aaronson07}, self-testing \cite{sf1,sf2,sf3,rac}, quantum randomness certification \cite{rangen}, quantum key distribution \cite{qkd}, studies of no-signaling resources \cite{Grudka} to characterising quantum mechanics from information-theoretic principles \cite{infoth}. Experimental demonstrations of RAC protocols have also been reported \cite{exprac1,exprac2}.
	
In the present study, we consider RAC using classical communications assisted with shared quantum correlations. In reality, it is experimentally difficult to create any quantum correlation. Moreover, environmental interactions unavoidably degrade the efficacy of any quantum correlation. To cope with these, one can recycle a single copy of any quantum resource multiple times. Furthermore, this also indicates how much quantumness in a correlation is preserved even after few cycles of local operations. Historically, this issue was first addressed by Silva \textit{et al.} \cite{sygp}, where two spatially separated spin-$\frac{1}{2}$ particles were assumed to be shared between a single Alice and multiple independent Bobs. In this scenario, the maximum number of Bobs was deduced \cite{sygp,majumdar,exp1,exp2,das,colbeck} that can demonstrate Bell nonlocality \cite{CHSH}.  
 This idea of sharing quantum correlations by multiple sequential observers has been extended in different contexts as well \cite{roykumar,sas,shenoy,Choi2020,bera,Foletto,malnew,saunak,rennew,expnew,pcon,akpan,balnew,Saha,Maity,Gupta}. The applications of sequential sharing of quantum correlations in different information processing tasks have also been demonstrated  \cite{rac,ran,cc,appln1,appln2,sroy,appln4,appln5}. In all these studies, multiple observers performing sequential measurements on only one qubit have been considered, whereas the present study contemplates multiple observers performing sequential measurements on each of the two qubits. This is a more general and practical scenario for re-utilizing quantum correlations in commercial quantum technologies.

In particular, we focus on recycling a single quantum resource in  sequentially carrying out RAC tasks multiple times. Here, we consider the scenario where a two-qubit state is shared between two spatially separated wings. Multiple independent Alices (say, Alice$^1$, Alice$^2$, Alice$^3$, $\cdots$) and multiple independent Bobs (say, Bob$^1$, Bob$^2$, Bob$^3$, $\cdots$) act sequentially on the first and second qubit respectively with unbiased inputs. At first, Alice$^1$-Bob$^1$ executes the RAC task with the initially shared two-qubit state. Afterwards, Alice$^1$ passes her qubit to Alice$^2$ and Bob$^1$ passes his qubit to Bob$^2$. Next,  Alice$^2$-Bob$^2$ also passes the two qubits to  Alice$^3$-Bob$^3$ after performing the RAC task and so on. 

In the above scenario, we show that unbounded pairs of Alice-Bob (i.e., Alice$^1$-Bob$^1$, Alice$^2$-Bob$^2$, $\cdots$) can gain quantum advantage in executing RAC tasks. Specifically, we demonstrate that the above result holds  1) when all pairs always perform some particular $2 \rightarrow 1$ RAC, 2) when all pairs always perform some particular $3 \rightarrow 1$ RAC task, 3) when each of the pairs always performs either a $2 \rightarrow 1$ RAC or a $3 \rightarrow 1$ RAC independent of other pairs, 4) when each pair performs a $2 \rightarrow 1$ RAC and a $3 \rightarrow 1$ RAC with different probabilities independent of other pairs. While comparing the classical and quantum strategies to demonstrate quantum advantage, we restrict the amount of shared classical bits to be equal to the amount of shared quantum bits. This constraint is quite natural in the sense that classical bits, similar to qubits, are expensive resources \cite{srac1,conbits1,sus,srac2}. Since, the aforementioned scenario involves two qubits, quantum strategies  are compared with the classical ones assisted with two bits from a common source. 


The rest of the paper is arranged as follows. In Section \ref{s2} we review the $2\rightarrow 1$ and $3\rightarrow 1$ RAC protocols assisted with classical communication and a two-qubit state. The scenario considered by us and the main results are presented in Section \ref{s3}. Finally, we conclude with a short discussion in Section \ref{s4}.

\section{$2\rightarrow 1$ and $3\rightarrow 1$ RAC protocols assisted with classical communication and a two-qubit state}\label{s2}
 Let us now describe the  $n\rightarrow 1$ (with $n \in \{2, 3\}$) RAC protocol using limited classical communication and shared two-qubit state. At first, Alice is given a string of $n$ bits $x = (x_0, x_1, \cdots, x_{n-1})$  chosen randomly from a uniform distribution with $x_i \in \lbrace 0,1 \rbrace$ for all $i \in \{0, 1, \cdots, n-1\}$. Next, depending on the input bit string, Alice performs one of the $2^n$ dichotomic measurements denoted by $A_{x_0 x_1 \cdots x_{n-1}}$ on her qubit. The outcome of the measurement $A_{x_0 x_1 \cdots x_{n-1}}$ is denoted by $a_{x_0 x_1 \cdots x_{n-1}} \in \{0, 1\}$. Alice then communicates the outcome of her measurement to Bob with one bit of information. Next, Bob tries to guess one of the $n$ bits $x_y$ (with $y \in \{0, 1, \cdots, n-1\}$) given to Alice (in each run $y$ is chosen randomly). For this purpose,   Bob performs one of the $n$ dichotomic measurements denoted by $B_{y}$ on his qubit. The outcome of the measurement $B_{y}$ is denoted by $b_{y} \in \{0, 1\}$. Finally, Bob's guess is given by $a_{x_0 x_1 \cdots x_{n-1}} \oplus b_y$. Hence, the RAC task will be successful, i.e., Bob's guess will be correct if and only if $a_{x_0 x_1 \cdots x_{n-1}} \oplus b_y = x_y$.

	In the present study, we will quantify the efficacy of the RAC protocol by minimum success probability defined as,
	\begin{align}
	P_{\texttt{Min}}^{n\rightarrow 1} & = \min\limits_{x_0, x_1, \cdots, x_{n-1}, y} \, \, P(a_{x_0 x_1 \cdots x_{n-1}} \oplus b_y = x_y).
	\label{minsuc}
	\end{align} 
	
\section{Results} \label{s3}
	
\begin{figure}
	{\color{white}
    \centering
    \includegraphics[width=250px,height=200px]{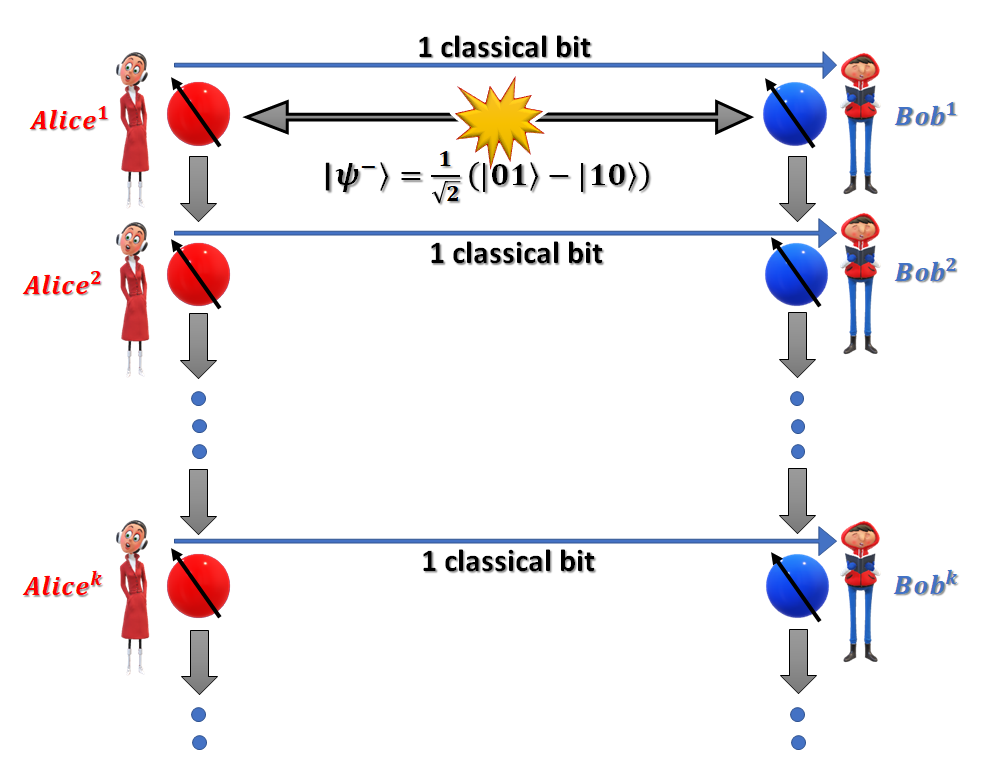}}
    \caption{Scenario for performing $n \rightarrow 1$ RAC task with multiple pairs of observers sequentially.}
    \label{fig1} 
\end{figure}

We consider a scenario involving multiple independent Alices and multiple independent Bobs as described in Fig. \ref{fig1}. Alice$^1$-Bob$^1$ initially shares one pair of qubit in the singlet state, $|\psi^{-}\rangle = \frac{1}{\sqrt{2}}(|01\rangle - |10\rangle)$. This first pair performs the aforementioned RAC task and then, Alice$^1$, Bob$^1$ pass their particles to Alice$^2$, Bob$^2$ respectively. Alice$^2$, Bob$^2$ also pass their particles to Alice$^3$, Bob$^3$ respectively after executing the RAC. In this way, the process continues. Note that each of the observers act with unbiased inputs. Here we want to find out how many pairs of Alice and Bob can exhibit quantum advantage. If any pair performs projective measurements, it will disturb the state maximally  and the next pair may not get any quantum advantage. Hence, in order to continue the above sequential RAC task with multiple pairs of Alice-Bob, we consider weak measurements by all pairs. We should choose the weak measurement formalism in such a way that the disturbance due to this measurement is minimized for any given amount of information gain \cite{sygp}. One such example is unsharp measurement (a particular class of Positive Operator-Valued Measure or POVM) \cite{pb2} with generalized von Neumann-L\"{u}ders state-transformation rule \cite{majumdar,sas}. 

In the present paper, we consider two particular RAC tasks. The first one is the $2 \rightarrow 1$ RAC task assisted with two (quantum or classical) bits, shared from a common source and having maximally mixed marginal  at the receiver's end. In the classical strategy, a source produces two correlated bits which are shared by Alice and Bob. The two binary values $0$ and $1$ of Bob's bit are equiprobable. Consequently, Alice's encoding and Bob's decoding strategies are now assisted with these bits. The minimum success probability of such a classical  RAC  task is always less than or equal to $\frac{1}{2}$ \cite{srac1}. In case of quantum strategy, two-qubit states with maximally mixed marginal at Bob's end can only be shared in the context of this task and  $P^{2 \rightarrow 1}_{\texttt{Min}} > \frac{1}{2}$ implies quantum advantage.

Another RAC task that we consider is the $3 \rightarrow 1$ RAC task assisted with two (quantum or classical) bits shared from a common source. There is no restriction on the marginals of the shared bits in this case. For classical strategies, the minimum success probability is always less than or equal to $\frac{1}{2}$ \cite{srac1}. Hence, $P^{3 \rightarrow 1}_{\texttt{Min}} > \frac{1}{2}$ ensures quantum advantage.

Suppose the pair Alice$^k$-Bob$^k$ for arbitrary $k \in \{1, 2, \cdots \}$ performs the above RAC using the shared Bell-diagonal state,
\begin{equation}
	 \rho^k_{AB}=\frac{1}{4} \left(\mathbf{I}_4 + \sum_{i=1}^{3} t^k_{ii} \, \sigma_{i} \otimes  \sigma_{i} \right),   
	 \label{belldiam}
	\end{equation} 
	where  $(t^k_{uu})^2 \geq (t^k_{vv})^2 \geq (t^k_{ww})^2$ for an arbitrary choice of $u \neq v \neq w \in \{1, 2, 3\}$; $\sigma_i$ with $i = 1, 2, 3$ are the three Pauli matrices.

Next, let us present the encoding-decoding strategies adopted by the pair Alice$^k$-Bob$^k$. In case of the $2 \rightarrow 1$ RAC, Alice$^k$ performs one of the four POVMs denoted by $A^k_{x_0 x_1} \equiv \{E^{k,0}_{x_0 x_1}, E^{k,1}_{x_0 x_1} \}$ with $(x_0, x_1) \in \{(00), (01), (10), (11)\}$, where
	\begin{equation}\label{unsharpA}
	E^{k, \, a^k_{x_0 x_1}}_{x_0 x_1} =\frac{1}{2}\left[ \mathbf{I}_{2}+ \lambda^k \, (-1)^{a^k_{x_0 x_1}} \, \, \left(\hat{u}^k_{x_0 x_1} \cdot \vec{\sigma} \right)   \right].
	\end{equation} 
Bob$^k$ performs one of the two POVMs denoted by $B^k_{y} \equiv \{E^{k,0}_{y}, E^{k,1}_{y} \}$ with $y \in \{0,1\}$, where
	\begin{equation}\label{unsharpB}
	E^{k, \, b^k_{y}}_{y} =\frac{1}{2}\left[ \mathbf{I}_{2}+ \eta^k \, (-1)^{b^k_{y}} \, \left(\hat{v}^k_{y} \cdot \vec{\sigma} \right) \right].
	\end{equation} 
	Here $\lambda^k$, $\eta^k$ $\in$ $(0, 1]$ are the sharpness parameters; $\vec{\sigma} = (\sigma_1, \sigma_2, \sigma_3)$; $a^k_{x_0 x_1}, \, b^k_{y} \in \{0, 1\}$ denote the outcomes of the POVMs $A^k_{x_0 x_1}$ performed by Alice$^k$ and $B^k_{y}$ performed by Bob$^k$ respectively. The unit vectors $\hat{u}^k_{x_0 x_1}$  and $\hat{v}^k_{y}$ are given by,
	\begin{alignat}{1}
	&\hat{u}^k_{x_0 x_1} = \left(\dfrac{ (-1)^{x_0} t^k_{11}}{\sqrt{(t^k_{11})^2+ (t^k_{22})^2}}, \dfrac{(-1)^{x_1} t^k_{22}}{\sqrt{(t^k_{11})^2+ (t^k_{22})^2}},0 \right), \label{encod}\\
	&\hat{v}^k_{0} =   \left(  1, 0, 0 \right), \quad \quad \quad \quad \quad \quad \quad \quad \quad \hat{v}^k_{1} =   \left(0, 1, 0 \right).	\label{decod}
    \end{alignat}
    
    On the other hand, for executing the aforementioned $3 \rightarrow 1$ RAC, Alice$^k$ performs one of the eight possible POVMs denoted by $A^k_{x_0 x_1 x_2} \equiv \{E^{k,0}_{x_0 x_1 x_2}, E^{k,1}_{x_0 x_1 x_2} \}$ with $x_i \in \lbrace 0,1 \rbrace$ for all $i \in \{0, 1, 2\}$, where
	\begin{equation}\label{unsharpA2}
	E^{k, \, a^k_{x_0 x_1 x_2}}_{x_0 x_1 x_2} =\frac{1}{2}\left[ \mathbf{I}_{2}+ \lambda^k \, (-1)^{a^k_{x_0 x_1 x_2}} \, \, \left(\hat{u}^k_{x_0 x_1 x_2} \cdot \vec{\sigma} \right)   \right].
	\end{equation} 
Bob$^k$ performs one of the three POVMs denoted by $B^k_{y} \equiv \{E^{k,0}_{y}, E^{k,1}_{y} \}$ with $y \in \{0,1,2\}$, where
	\begin{equation}\label{unsharpB2}
	E^{k, \, b^k_{y}}_{y} =\frac{1}{2}\left[ \mathbf{I}_{2}+ \eta^k \, (-1)^{b^k_{y}} \, \left(\hat{v}^k_{y} \cdot \vec{\sigma} \right) \right].
	\end{equation} 
	We choose the unit vectors $\hat{u}^k_{x_0 x_1 x_2} = \dfrac{\vec{u}^k_{x_0 x_1 x_2}}{|\vec{u}^k_{x_0 x_1 x_2}|}$  and $\hat{v}^k_{y}$ as follows,
	\begin{alignat}{1}
	&\vec{u}^k_{x_0 x_1 x_2} = \Big((-1)^{x_0} t^k_{11}, \, (-1)^{x_1} t^k_{22}, \, (-1)^{x_2} t^k_{33} \Big), \label{encod2}\\
	&\hat{v}^k_{0} =   \left( 1, 0, 0 \right), \quad \hat{v}^k_{1} =   \left( 0, 1, 0 \right), \quad \hat{v}^k_{2} =   \left( 0, 0, 1 \right).	\label{decod2}
    \end{alignat}

With these, we can present the following lemma (for proof, see Appendix \ref{appendix1}), which will be useful for probing the main result,
\begin{Lemma}
Let Alice$^k$-Bob$^k$ performs the $n \rightarrow 1$ RAC task (where $n=2$ or $n=3$) with a two-qubit Bell-diagonal state (\ref{belldiam}) using the above unsharp measurements. Then the pair achieves minimum success probability strictly greater than $\frac{1}{2}$ if $\min\limits_{i \leq n} \left[ (t^k_{ii})^2 \right] \neq 0$.
\label{lemma1}
\end{Lemma}
Next, we want to find out the  post-measurement state $\rho^{k+1}_{AB}$  received, on average, by Alice$^{k+1}$-Bob$^{k+1}$ from Alice$^{k}$-Bob$^{k}$. When Alice$^{k}$-Bob$^{k}$ performs the $2 \rightarrow 1$ RAC, following the generalized von Neumann-L\"{u}der's transformation rule, we have (see Appendix \ref{appendix2})
		\begin{align}
	\rho^{k+1}_{AB} &= \frac{1}{8}  \sum_{x_0, x_1 , y=0}^{1} \Bigg[ \sum_{a^k_{x_0 x_1}, b^k_y=0}^{1}  \Bigg( \sqrt{E^{k, \, a^k_{x_0 x_1}}_{x_0 x_1}} \otimes \sqrt{E^{k, \, b^k_y}_{y}} \Bigg) \nonumber \\
	& \quad \quad \quad \quad \quad \quad \quad \quad \rho_{AB}^k \Bigg( \sqrt{E^{k, \, a^k_{x_0 x_1}}_{x_0 x_1}} \otimes \sqrt{E^{k, \, b^k_y}_{y}} \Bigg)^{\dagger} \Bigg] \nonumber \\
	& =\frac{1}{4} \left(\mathbf{I}_4 +\sum_{i=1}^{3}t^{k+1}_{ii} \, \sigma_{i} \otimes  \sigma_{i} \right).
	\label{postappb}
	\end{align} 
The average is taken since we have assumed that multiple Alices or multiple Bobs act independently of each other. Here, we have also used the assumption that Alice$^k$ and Bob$^k$ perform measurements with unbiased inputs. Similarly, when Alice$^{k}$-Bob$^{k}$ performs the $3 \rightarrow 1$ RAC, it is observed that the average post-measurement state $\rho^{k+1}_{AB}$  received by Alice$^{k+1}$-Bob$^{k+1}$  has the Bell-diagonal form (\ref{postappb}) (see Appendix \ref{appendix3} for details).

Moreover, when Alice$^k$-Bob$^k$ performs the $n \rightarrow 1$ RAC task (where $n=2$ or $n=3$) with the state (\ref{belldiam}), it can be shown that $\min\limits_{i \leq n} \left[ (t^{k+1}_{ii})^2 \right] \neq 0$ if $\min\limits_{i \leq n} \left[ (t^k_{ii})^2 \right] \neq 0$ (For details, see Appendix \ref{appendix2} and Appendix \ref{appendix3}).

Now, consider that the same $n \rightarrow 1$ RAC (i.e., either the $2 \rightarrow 1$ or the $3 \rightarrow 1$ RAC) is performed by each of the pairs. In such scenario, combining the above results, we can present the following: if Alice$^1$-Bob$^1$ initially shares the singlet state, then this pair achieves $P^{n \rightarrow 1}_{\texttt{Min}} > \frac{1}{2}$ (with $n=2$ or $n=3$) using the aforementioned unsharp measurements. Moreover, the average post-measurement state $\rho^2_{AB}$ received by Alice$^2$-Bob$^2$ is the Bell-diagonal state (\ref{belldiam}) with $k=2$ and $\min\limits_{i \leq n} \left[ (t^{2}_{ii})^2 \right] \neq 0$. Hence,  Alice$^2$-Bob$^2$ also achieves $P^{n \rightarrow 1}_{\texttt{Min}} > \frac{1}{2}$. Subsequently, Alice$^3$-Bob$^3$ receives  the Bell-diagonal state (\ref{belldiam}) with $k=3$ and $\min\limits_{i \leq n} \left[ (t^{3}_{ii})^2 \right] \neq 0$ and  exhibits $P^{n \rightarrow 1}_{\texttt{Min}} > \frac{1}{2}$ as well. This process continues for arbitrarily many  pairs. Therefore, we can present the following theorem,

\begin{thm}
Unbounded pairs of Alice and Bob can demonstrate quantum advantage either in $2 \rightarrow 1$ RAC task assisted with two bits shared from a common source and having maximally mixed marginal  at the receiver's end, or in  $3 \rightarrow 1$ RAC task assisted with two correlated bits.
\label{mintheo}
\end{thm}





Importantly, the statements of Theorem \ref{mintheo} hold for all values of $\lambda^k$ $\in$ $(0, 1]$ and $\eta^k$ $\in$ $(0,1]$ for all possible $k$ $\in$ $\{1, 2, \cdots\}$.  Moreover, for the aforementioned $n \rightarrow 1$ RAC with $n=2$ or $n=3$, starting with any Bell-diagonal two-qubit (entangled or separable) state given by Eq.(\ref{belldiam}) with $k=1$ and $\min\limits_{i \leq n} \left[ (t^{1}_{ii})^2 \right] \neq 0$, one gets the same result  as stated in Theorem \ref{mintheo}. Hence, the following  corollary can be stated,

\begin{cor}
Unbounded pairs of Alice and Bob can exhibit quantum advantage in some particular $n \rightarrow 1$ RAC task (with $n = 2$ or $n=3$) even when each of the observers performs suitable projective measurements and the initially shared two-qubit state belongs to a particular subset of separable states.
\label{corollary1}
\end{cor}

When Alice$^1$-Bob$^1$ initially shares the singlet state and performs the aforementioned $n \rightarrow 1$ RAC (where $n=2$ or $n=3$) using the   measurements described earlier with $\lambda^1= \eta^1 = 1$ (i.e., projective measurements), then this pair achieves $P^{n \rightarrow 1}_{\texttt{Min}} = \frac{1}{2} \left( 1 + \frac{1}{\sqrt{n}}\right)$. This is the maximum permissible value of  $P^{n \rightarrow 1}_{\texttt{Min}}$ with quantum resources \cite{erac}. In this case also, the residual quantum correlation in the average post-measurement state is sufficient for demonstrating quantum advantage in the  $n \rightarrow 1$ RAC by  unbounded pairs of Alice and Bob. Hence, a single pair of qubits can be utilized indefinitely to gain quantum advantage in some particular RAC even when the optimal quantum advantage is exhibited in the first round.

\textbf{Remark:} We observe that when an arbitrary pair gains a large amount of quantum advantage, then only few numbers of subsequent pairs will get `significant' quantum advantage. On the other hand, when a pair gets a small amount of quantum advantage, a larger number of subsequent pairs can achieve `significant' quantum advantage. Here, `significant' quantum advantage implies that  $\left(P_{\texttt{Min}}^{n \rightarrow 1} - \frac{1}{2} \right)$ is positive and large enough to be detected in a real experiment. Hence, there may exist a trade-off relation between the amount of quantum advantage gained by an arbitrary pair and the number of subsequent pairs exhibiting considerable amount of quantum advantage. Moreover, either of these two quantities can be increased at the expense of the other by suitably choosing the sharpness parameters of the measurements (See Appendix \ref{appendix4}). In practical scenario, a large but finite number of sequential pairs of observers may be required to perform some communication tasks with only one pair of qubits. The number of sequential pairs required to exhibit quantum advantage  depends on the particular context under consideration and that can be realized by fine-tuning the unsharpness of the measurements.     

Next, we consider a more general scenario where an arbitrary pair Alice$^k$-Bob$^k$ performs the aforementioned $2 \rightarrow 1$ RAC task with probability $p_k$ and the aforementioned $3 \rightarrow 1$ RAC with probability $(1-p_k)$, where $0 \leq p_k \leq 1$. For example, Alice$^k$ and Bob$^k$ can fix the task to be performed in each experimental run prior to the initiation of sequential RAC and, during the execution of sequential RAC, they perform the two different tasks accordingly. This type of scenario is particularly relevant when a sequence of RAC tasks is implemented as an intermediate step in commercial quantum computation. In such cases, different tasks may be required to be performed by the same pair of particles in different steps depending on the choices of users. In this scenario, if a singlet state or any Bell-diagonal two-qubit (entangled or separable) state given by Eq.(\ref{belldiam}) with $k=1$ and $\min \left[(t^{1}_{11})^2, (t^{1}_{22})^2, (t^{1}_{33})^2 \right] \neq 0$ is initially shared, then the following result is attained (see Appendix \ref{appendix5} for details),

\begin{cor}
Unbounded pairs of Alice and Bob can demonstrate quantum advantage when an arbitrary pair Alice$^k$-Bob$^k$ performs a $2 \rightarrow 1$ RAC (assisted with two correlated bits with maximally mixed marginal at the receiver's end) with probability $p_k$ and a $3 \rightarrow 1$ RAC (assisted with two bits shared from a common source) with probability $1-p_k$ independent of other pairs.
\label{mintheo3}
\end{cor}

When Alice$^k$-Bob$^k$ performs projective measurements and $p_k =1$ (i.e., performs $2 \rightarrow 1$ RAC with certainty), then the condition: $\min \left[(t^{x}_{11})^2, (t^{x}_{22})^2, (t^{x}_{33})^2 \right] \neq 0$ will not be satisfied for the average post-measurement state received by all subsequent pairs (i.e., for all $x \in \{k+1, k+2, \cdots \}$). Hence, all these pairs will not achieve quantum advantage in $3 \rightarrow 1$ RAC. Hence, only under unsharp measurements (with the sharpness parameters being strictly less than $1$), we can state the following corollary (see Appendix \ref{appendix5} for details),
\begin{cor}
Unbounded pairs of Alice and Bob can demonstrate quantum advantage when an arbitrary pair Alice$^k$-Bob$^k$ performs a $2 \rightarrow 1$ RAC with certainty and another arbitrary pair Alice$^{\tilde{k}}$-Bob$^{\tilde{k}}$ performs a  $3 \rightarrow 1$ RAC with certainty for all choices of $k \neq \tilde{k} \in \{1, 2, \cdots\}$.
\label{cor2}
\end{cor}

\section{Conclusions} \label{s4}
Here we have considered a scenario involving multiple independent pairs of Alice and Bob sharing a single pair of qubits and performing some particular $2 \rightarrow 1$ and $3 \rightarrow 1$ RAC tasks  with unbiased inputs sequentially. In this scenario, we have shown that  unbounded pairs can gain quantum advantage  even when all observers perform projective measurements. These results address the issue of recycling a single copy of a quantum resource in performing information processing tasks multiple times sequentially. This is of utmost importance since, in reality, preparing quantum correlations and preserving them against inevitable environmental interactions are difficult. 


Our results point out that quantum correlations present in separable states \cite{discord} can be preserved indefinitely in spite of utilizing it in each step. Furthermore, weak measurements are not necessary for this purpose; suitable projective measurements can serve for this. Note that this is not the case for entanglement or Bell-nonlocality. Hence, these results signify one fundamental difference between the quantum correlations present in entanglement and that present in separable states: the first one is destroyed only after one cycle of projective measurements while the second one is retained even after infinite cycles. The advantage of quantum information processing tasks assisted with separable states \cite{srac1,srac2} is thus pointed out by our present study.  In fact, our results open up the possibility of implementing unbounded sequence of any task, for which quantum advantage can be demonstrated even using separable states (say, for example, remote state preparation \cite{rsp}),  with only one pair of qubits.

There exists a complementarity between the question addressed here and the one-way communication complexity problem \cite{cce1,cce2}. In one-way communication complexity problem, Alice and Bob are  given inputs $x \in \left\lbrace 0,1 \right\rbrace^{n}$ and $y \in \left\lbrace 0,1 \right\rbrace^{m}$ respectively. The goal for Bob is to calculate a binary function $f(x, y)$. Alice is allowed to send limited classical communications to Bob. This game can be thought as a number of parallel RACs taking place simultaneously. The main goal of any communication complexity problem is to minimize the amount of classical communication. However, there is no restriction on the shared entanglement. On the contrary, the present study is aimed to reduce the amount of shared correlation, but does not focus on reducing the number of communicating bits.

Recently, measurement protocols have been proposed to demonstrate arbitrary many Bell-CHSH inequality \cite{CHSH} violations with various independent Bobs and a single Alice using unbiased inputs when a pure entangled two-qubit  state is initially shared  \cite{colbeck}. The result, however, requires arbitrarily high precision engineering for the measurement apparatus and, hence, is too strenuous to implement in a reality. On the other hand, the unsharp measurements chosen in the present study can be realized in photonic systems based on the techniques adopted in \cite{appln4,appln5}. Moreover, our results are valid for any range of sharpness parameters and do not require any entanglement. Hence, for experimental implementation of large sequence of detecting quantum correlation with a single two-qubit state, our results are less demanding.


To the best of our knowledge, this study points out for the first time that there exist some communication tasks in which unbounded pairs of observers can exhibit quantum supremacy even if a single quantum resource is used. Finding out different communication tasks with the above feature merits further investigation. Next, it is worth to fully characterize the set of two-qubit states for which Theorem \ref{mintheo} holds. It is also interesting to find out whether there exists any two-qubit state for which weak measurements are necessary for satisfying Theorem \ref{mintheo}. \\ 

\section*{Acknowledgements}
DD acknowledges  fruitful  discussions  with Somshubhro Bandyopadhyay, Manik Banik and  Debashis Saha. DD acknowledges  Science and Engineering Research Board (SERB), Government of India for financial support through National Post Doctoral Fellowship (File No.: PDF/2020/001358).  AG acknowledges Bose Institute, Kolkata for financial support. SK thanks the Department of Science and Technology (DST), Government of India for the financial assistance through the QuEST project

		\newpage
		
		\appendix

\begin{widetext}
			\section{Proof of Lemma \ref{lemma1}}\label{appendix1}

{\bf For the $2 \rightarrow 1$ RAC:}	Let Alice$^k$-Bob$^k$ shares the following Bell-diagonal two-qubit state,
	\begin{equation}
	 \rho^k_{AB}=\frac{1}{4}(\mathbf{I}_4 +\sum_{i=1}^{3}t^k_{ii} \, \sigma_{i }\otimes  \sigma_{i}), 
	 \label{belldiaa4}
	\end{equation} 
	with $(t^k_{uu})^2 \geq (t^k_{vv})^2 \geq (t^k_{ww})^2$ for an arbitrary choice of $u \neq v \neq w \in \{1, 2, 3\}$. 
	
	Alice$^k$ and Bob$^k$ perform the $2 \rightarrow 1$ RAC task contingent upon using the unsharp measurements mentioned in Eqs.(\ref{unsharpA},\ref{unsharpB},\ref{encod},\ref{decod}). Next, let us compute the expression for a typical guessing probability $P(a^k_{x_0 x_1} \oplus b^k_y = x_y)$. Using Born's rule, one can write that
	\begin{align}
	P(a^k_{x_0 x_1} \oplus b^k_y = x_y)
	&= \sum_{z=0}^{1}P\left(a^k_{x_0 x_1} =z , b^k_y = |x_y-z| \, \, \Big| \, \, A^k_{x_0 x_1}, B^k_y \right),
	\label{supmat2}
	\end{align}
	where $P\left(a^k_{x_0 x_1} =z , b^k_y = |x_y-z| \, \, \Big| \, \, A^k_{x_0 x_1}, B^k_y \right)$ denotes the joint probability with which Alice$^k$ and Bob$^k$ get the outcomes $z$ and $|x_y-z|$ contingent upon performing the measurements $A^k_{x_0 x_1}$ and $B^k_y$ respectively. From Eq.(\ref{supmat2}), we have the following,
\begin{align}
	P(a^k_{x_0 x_1} \oplus b^k_y = x_y)
	&= \sum_{z=0}^{1} \text{Tr}\left[ \rho^k_{AB} \left(E^{k, \, z}_{x_0 x_1} \otimes E^{k, \, |x_y-z|}_{y} \right)\right] \nonumber \\
	&=\frac{1}{2}\left[ 1+ (-1)^{x_y} \, \lambda^k \eta^k \, \left(t_{(y+1) \, (y+1)} \right) \,  \left(\hat{u}_{x_0 x_1}^{k} \cdot \hat{v}_{y}^{k} \right) \right],
	\label{supmat3}
	\end{align} 
where $\hat{u}^k_{x_0 x_1}$	and $\hat{v}^k_{y}$ are mentioned  in Eqs.(\ref{encod},\ref{decod}). Now, a straightforward calculation leads to the following, 
\begin{subequations}
    \begin{equation}
    P(a^k_{00}\oplus b^k_{0}=0)=P(a^k_{01}\oplus b^k_{0}=0) = P(a^k_{10}\oplus b^k_{0}=1) = P(a^k_{11}\oplus b^k_{0}=1) = \frac{1}{2}\left[ 1+\lambda^{k}\eta^{k}\frac{(t^k_{11})^2}{\sqrt{(t^k_{11})^{2}+(t^k_{22})^{2}}} \right],
    \end{equation}
    \begin{equation}
    P(a^k_{00}\oplus b^k_{1}=0)= P(a^k_{01}\oplus b^k_{1}=1)=P(a^k_{10}\oplus b^k_{1}=0)=P(a^k_{11}\oplus b^k_{1}=1)=\frac{1}{2}\left[ 1+\lambda^{k}\eta^{k}\frac{(t^k_{22})^2}{\sqrt{(t^k_{11})^{2}+(t^k_{22})^{2}}} \right].
    \end{equation}
    \end{subequations}
Hence, the minimum success probability is given by,
    \begin{equation}
        P_{\texttt{Min}}^{2\rightarrow 1}=\frac{1}{2}\left[ 1+\lambda^{k}\eta^{k}\frac{\min \left[(t^k_{11})^2, (t^k_{22})^2 \right]}{\sqrt{(t^k_{11})^{2}+(t^k_{22})^{2}}} \right].
    \end{equation}
Thus, for all values of $\lambda^{k}$ $\in$ $(0, 1]$ and $\eta^{k}$ $\in$ $(0, 1]$, $P_{\texttt{Min}}^{2\rightarrow 1}>\frac{1}{2}$ if $\min \left[(t^{k}_{11})^2, (t^{k}_{22})^2 \right] \neq 0$. \\\\

{\bf For the $3 \rightarrow 1$ RAC:} 	Let Alice$^k$-Bob$^k$ performs  the $3 \rightarrow 1$ RAC task  using the shared Bell-diagonal state given by Eq. (\ref{belldiaa4}).  Alice$^k$ and Bob$^k$ perform the  unsharp measurements given by Eqs.(\ref{unsharpA2},\ref{unsharpB2},\ref{encod2},\ref{decod2}) of the main paper. The expression for a typical guessing probability $P(a^k_{x_0 x_1 x_2} \oplus b^k_y = x_y)$ can be calculated using Born's rule as follows
\begin{align}
	P(a^k_{x_0 x_1 x_2} \oplus b^k_y = x_y)
	&= \sum_{z=0}^{1} \text{Tr}\left[ \rho^k_{AB} \left(E^{k, \, z}_{x_0 x_1 x_2} \otimes E^{k, \, |x_y-z|}_{y} \right)\right] \nonumber \\
	&=\frac{1}{2}\left[ 1+ (-1)^{x_y} \, \lambda^k \eta^k \, \left(t_{(y+1) \, (y+1)} \right) \,  \left(\hat{u}_{x_0 x_1 x_2}^{k} \cdot \hat{v}_{y}^{k} \right) \right],
	\label{supmat3}
	\end{align} 
where $\hat{u}^k_{x_0 x_1 x_2}$	and $\hat{v}^k_{y}$ are mentioned earlier in Eqs.(\ref{encod2},\ref{decod2}) of the main paper. Hence, we have the following, 
    \begin{align}
    P(a^k_{000}\oplus b^k_{0}=0)=P(a^k_{001}\oplus b^k_{0}=0) = P(a^k_{010}\oplus b^k_{0}=0) = P(a^k_{011}\oplus b^k_{0}=0) = \frac{1}{2}\left[ 1+\lambda^{k}\eta^{k}\frac{(t^k_{11})^2}{\sqrt{(t^k_{11})^{2}+(t^k_{22})^{2} + (t^k_{33})^{2}}} \right], \nonumber \\
    P(a^k_{100}\oplus b^k_{0}=1)=P(a^k_{101}\oplus b^k_{0}=1) = P(a^k_{110}\oplus b^k_{0}=1) = P(a^k_{111}\oplus b^k_{0}=1) = \frac{1}{2}\left[ 1+\lambda^{k}\eta^{k}\frac{(t^k_{11})^2}{\sqrt{(t^k_{11})^{2}+(t^k_{22})^{2} + (t^k_{33})^{2}}} \right], \nonumber \\
    P(a^k_{000}\oplus b^k_{1}=0)=P(a^k_{001}\oplus b^k_{1}=0) = P(a^k_{010}\oplus b^k_{1}=1) = P(a^k_{011}\oplus b^k_{1}=1) = \frac{1}{2}\left[ 1+\lambda^{k}\eta^{k}\frac{(t^k_{22})^2}{\sqrt{(t^k_{11})^{2}+(t^k_{22})^{2} + (t^k_{33})^{2}}} \right], \nonumber \\
    P(a^k_{100}\oplus b^k_{1}=0)=P(a^k_{101}\oplus b^k_{1}=0) = P(a^k_{110}\oplus b^k_{1}=1) = P(a^k_{111}\oplus b^k_{1}=1) = \frac{1}{2}\left[ 1+\lambda^{k}\eta^{k}\frac{(t^k_{22})^2}{\sqrt{(t^k_{11})^{2}+(t^k_{22})^{2}+ (t^k_{33})^{2}}} \right], \nonumber \\
    P(a^k_{000}\oplus b^k_{2}=0)=P(a^k_{001}\oplus b^k_{2}=1) = P(a^k_{010}\oplus b^k_{2}=0) = P(a^k_{011}\oplus b^k_{2}=1) = \frac{1}{2}\left[ 1+\lambda^{k}\eta^{k}\frac{(t^k_{33})^2}{\sqrt{(t^k_{11})^{2}+(t^k_{22})^{2} + (t^k_{33})^{2}}} \right], \nonumber \\
    P(a^k_{100}\oplus b^k_{2}=0)=P(a^k_{101}\oplus b^k_{2}=1) = P(a^k_{110}\oplus b^k_{2}=0) = P(a^k_{111}\oplus b^k_{2}=1) = \frac{1}{2}\left[ 1+\lambda^{k}\eta^{k}\frac{(t^k_{33})^2}{\sqrt{(t^k_{11})^{2}+(t^k_{22})^{2}+ (t^k_{33})^{2}}} \right].
    \end{align}
From the above equations, the minimum success probability is given by,
    \begin{equation}
        P_{\texttt{Min}}^{3\rightarrow 1}=\frac{1}{2}\left[ 1+\lambda^{k}\eta^{k}\frac{\min \left[(t^k_{11})^2, (t^k_{22})^2, (t^k_{33})^2 \right]}{\sqrt{(t^k_{11})^{2}+(t^k_{22})^{2}+ (t^k_{33})^{2}}} \right].
    \end{equation}
Therefore, for all values of $\lambda^{k}$ $\in$ $(0, 1]$ and $\eta^{k}$ $\in$ $(0, 1]$, $P_{\texttt{Min}}^{3\rightarrow 1}>\frac{1}{2}$ if $\min \left[(t^{k}_{11})^2, (t^{k}_{22})^2, (t^{k}_{33})^2 \right] \neq 0$. 

\section{Calculating $\rho^{k+1}_{AB}$ in case of $2 \rightarrow 1$ RAC}\label{appendix2}
	
	Let Alice$^k$-Bob$^k$ performs the $2 \rightarrow 1$ RAC task with the following Bell-diagonal two-qubit state,
	\begin{equation}
	 \rho^k_{AB}=\frac{1}{4}(\mathbf{I}_4 +\sum_{i=1}^{3}t^k_{ii} \, \sigma_{i} \otimes  \sigma_{i}),   
	 \label{belldia}
	\end{equation} 
	where  $(t^k_{uu})^2 \geq (t^k_{vv})^2 \geq (t^k_{ww})^2$ for an arbitrary choice of $u \neq v \neq w \in \{1, 2, 3\}$. 
	
	Alice$^k$ and Bob$^k$ perform the aforementioned unsharp measurements. The average post-measurement state $\rho_{AB}^{k+1}$ received by Alice$^{k+1}$-Bob$^{k+1}$ from Alice$^k$-Bob$^k$ can be obtained using the generalized von Neumann-L\"{u}der's transformation rule as follows, 
		\begin{align}
	\rho^{k+1}_{AB} = \frac{1}{8}  \sum_{x_0, x_1 , y=0}^{1} & \Bigg[ \sum_{a^k_{x_0 x_1}, b^k_y=0}^{1}  \Bigg( \sqrt{E^{k, \, a^k_{x_0 x_1}}_{x_0 x_1}} \otimes \sqrt{E^{k, \, b^k_y}_{y}} \Bigg) \rho^{k}_{AB} \Bigg( \sqrt{E^{k, \, a^k_{x_0 x_1}}_{x_0 x_1}} \otimes \sqrt{E^{k, \, b^k_y}_{y}} \Bigg)^{\dagger} \Bigg],
	\label{postappbsm}
	\end{align}   
	where
	\begin{equation}
	\sqrt{E^{k, \, a^k_{x_0 x_1}}_{x_0 x_1}} = \frac{\sqrt{1+\lambda^k}}{2 \sqrt{2}}\left[ \mathbf{I}_{2}+   (-1)^{a^k_{x_0 x_1}} \, \left(\hat{u}^k_{x_0 x_1} \cdot \vec{\sigma} \right) \right] + \frac{\sqrt{1-\lambda^k}}{2 \sqrt{2}}\left[ \mathbf{I}_{2}-  (-1)^{a^k_{x_0 x_1}} \, \left(\hat{u}^k_{x_0 x_1} \cdot \vec{\sigma} \right) \right],
	\label{sqrtalice}
	\end{equation} 
	and
	\begin{equation}
	\sqrt{E^{k, \, b^k_{y}}_{y}} =\frac{\sqrt{1+\eta^k}}{2 \sqrt{2}}\left[ \mathbf{I}_{2}+  (-1)^{b^k_{y}} \, \left(\hat{v}^k_{y} \cdot \vec{\sigma} \right) \right] + \frac{\sqrt{1-\eta^k}}{2 \sqrt{2}}\left[ \mathbf{I}_{2}-  (-1)^{b^k_{y}} \, \left(\hat{v}^k_{y} \cdot \vec{\sigma} \right) \right],
	\label{sqrtbob}
	\end{equation} 
with $\hat{u}^k_{x_0 x_1}$	and $\hat{v}^k_{y}$ being mentioned in Eqs.(\ref{encod},\ref{decod}). Using Eqs.(\ref{postappbsm}-\ref{sqrtbob}), one has the following, \begin{equation}
	 \rho^{k+1}_{AB}=\frac{1}{4}(\mathbf{I}_4 +\sum_{i=1}^{3} t^{k+1}_{ii} \, \sigma_{i }\otimes  \sigma_{i}),   
	 \label{belldia2}
\end{equation} 
with
\begin{align}
 t^{k+1}_{11} &= \frac{t^k_{11} \, \left[1+ \sqrt{1-(\eta^k)^2}\right] \left[(t^k_{11})^2 + (t^k_{22})^2 \sqrt{1- (\lambda^k)^2} \right]}{2 \left[(t^k_{11})^2 + (t^k_{22})^2\right]},   \nonumber \\
 t^{k+1}_{22} &= \frac{t^k_{22} \, \left[1+ \sqrt{1-(\eta^k)^2} \right]  \left[(t^k_{22})^2 + (t^k_{11})^2 \sqrt{1-(\lambda^k)^2}\right]}{2 \left[(t^k_{11})^2 + (t^k_{22})^2\right]},   \nonumber \\
 t^{k+1}_{33} &= t^k_{33} \, \sqrt{1-(\eta^k)^2} \sqrt{1-(\lambda^k)^2}.
 \label{tevsm2}
\end{align}

Hence, the average post-measurement state $\rho^{k+1}_{AB}$ is Bell-diagonal. Further, we have $\min \left[(t^{k+1}_{11})^2, (t^{k+1}_{22})^2 \right] \neq 0$ if $\min \left[(t^{k}_{11})^2, (t^{k}_{22})^2 \right] \neq 0$ for all possible values of $\lambda^{k}$ $\in$ $(0, 1]$ and $\eta^{k}$ $\in$ $(0, 1]$.

In particular, when $\lambda^k = \eta^k = 1$, we have
\begin{align}
 t^{k+1}_{11} &= \frac{(t^k_{11})^{3}}{2 \left[(t^k_{11})^2 + (t^k_{22})^2\right]},   \nonumber \\
 t^{k+1}_{22} &= \frac{(t^k_{22})^3}{2 \left[(t^k_{11})^2 + (t^k_{22})^2\right]},   \nonumber \\
 t^{k+1}_{33} &= 0.
\end{align}
Therefore, the state $\rho^{k+1}_{AB}$ remains to be Bell-diagonal with $\min \left[(t^{k+1}_{11})^2, (t^{k+1}_{22})^2 \right] \neq 0$ if $\min \left[(t^{k}_{11})^2, (t^{k}_{22})^2 \right] \neq 0$ even when Alice$^k$-Bob$^k$ performs projective measurements.


	\section{Calculating $\rho^{k+1}_{AB}$ in case of $3 \rightarrow 1$ RAC}\label{appendix3}
	
	Let Alice$^k$-Bob$^k$ performs the $3 \rightarrow 1$ RAC task with the following Bell-diagonal two-qubit state,
	\begin{equation}
	 \rho^k_{AB}=\frac{1}{4}(\mathbf{I}_4 +\sum_{i=1}^{3}t^k_{ii} \, \sigma_{i} \otimes  \sigma_{i}),   
	 \label{belldia2}
	\end{equation} 
	where  $(t^k_{uu})^2 \geq (t^k_{vv})^2 \geq (t^k_{ww})^2$ for an arbitrary choice of $u \neq v \neq w \in \{1, 2, 3\}$. 
	
	Alice$^k$ and Bob$^k$ perform the  unsharp measurements mentioned in Eqs.(\ref{unsharpA2},\ref{unsharpB2},\ref{encod2},\ref{decod2}). The average post-measurement state $\rho_{AB}^{k+1}$ received by Alice$^{k+1}$-Bob$^{k+1}$ from Alice$^k$-Bob$^k$ can be obtained using the generalized von Neumann-L\"{u}der's transformation rule and it is given by,  
		\begin{align}
	\rho^{k+1}_{AB} = \frac{1}{24}  \sum_{x_0, x_1 , x_2=0}^{1} \, \, \, \, \, \sum_{y=0}^{1} & \Bigg[ \sum_{a^k_{x_0 x_1 x_2}=0}^{1} \, \, \, \, \, \, \sum_{ b^k_y=0}^{1} \Bigg( \sqrt{E^{k, \, a^k_{x_0 x_1 x_2}}_{x_0 x_1 x_2}} \otimes \sqrt{E^{k, \, b^k_y}_{y}} \Bigg) \rho^{k}_{AB} \Bigg( \sqrt{E^{k, \, a^k_{x_0 x_1 x_2}}_{x_0 x_1 x_2}} \otimes \sqrt{E^{k, \, b^k_y}_{y}} \Bigg)^{\dagger} \Bigg],
	\label{postappb2}
	\end{align}   
	where
	\begin{equation}
	\sqrt{E^{k, \, a^k_{x_0 x_1 x_2}}_{x_0 x_1 x_2}} = \frac{\sqrt{1+\lambda^k}}{2 \sqrt{2}}\left[ \mathbf{I}_{2}+   (-1)^{a^k_{x_0 x_1 x_2}} \, \left(\hat{u}^k_{x_0 x_1 x_2} \cdot \vec{\sigma} \right) \right] + \frac{\sqrt{1-\lambda^k}}{2 \sqrt{2}}\left[ \mathbf{I}_{2}-  (-1)^{a^k_{x_0 x_1 x_2}} \, \left(\hat{u}^k_{x_0 x_1 x_2} \cdot \vec{\sigma} \right) \right],
	\label{sqrtalice2}
	\end{equation} 
	and
	\begin{equation}
	\sqrt{E^{k, \, b^k_{y}}_{y}} =\frac{\sqrt{1+\eta^k}}{2 \sqrt{2}}\left[ \mathbf{I}_{2}+  (-1)^{b^k_{y}} \, \left(\hat{v}^k_{y} \cdot \vec{\sigma} \right) \right] + \frac{\sqrt{1-\eta^k}}{2 \sqrt{2}}\left[ \mathbf{I}_{2}-  (-1)^{b^k_{y}} \, \left(\hat{v}^k_{y} \cdot \vec{\sigma} \right) \right]
	\label{sqrtbob2}
	\end{equation} 
with $\hat{u}^k_{x_0 x_1 x_2}$	and $\hat{v}^k_{y}$ are mentioned in Eq.(\ref{encod2}) and Eq.(\ref{decod2}) respectively. Using Eqs.(\ref{postappb2}-\ref{sqrtbob2}), we have the following, 
\begin{equation}
	 \rho^{k+1}_{AB}=\frac{1}{4}(\mathbf{I}_4 +\sum_{i=1}^{3} t^{k+1}_{ii} \, \sigma_{i }\otimes  \sigma_{i}),   
	 \label{belldia22}
\end{equation} 
with
\begin{align}
 t^{k+1}_{11} &= \frac{t^k_{11} \, \left[1+ 2 \sqrt{1-(\eta^k)^2}\right] \left[(t^k_{11})^2 + \left[ (t^k_{22})^2 + (t^k_{33})^2 \right] \sqrt{1- (\lambda^k)^2} \right]}{3 \left[(t^k_{11})^2 + (t^k_{22})^2 + (t^k_{33})^2\right]},   \nonumber \\
 t^{k+1}_{22} &= \frac{t^k_{22} \, \left[1+ 2 \sqrt{1-(\eta^k)^2}\right] \left[(t^k_{22})^2 + \left[ (t^k_{33})^2 + (t^k_{11})^2 \right] \sqrt{1- (\lambda^k)^2} \right]}{3 \left[(t^k_{11})^2 + (t^k_{22})^2 + (t^k_{33})^2\right]} ,  \nonumber \\
 t^{k+1}_{33} &= \frac{t^k_{33} \, \left[1+ 2 \sqrt{1-(\eta^k)^2}\right] \left[(t^k_{33})^2 + \left[ (t^k_{11})^2 + (t^k_{22})^2 \right] \sqrt{1- (\lambda^k)^2} \right]}{3 \left[(t^k_{11})^2 + (t^k_{22})^2 + (t^k_{33})^2\right]} .
 \label{tevsm4}
\end{align}

Hence, the average post-measurement state $\rho^{k+1}_{AB}$ is Bell-diagonal. Furthermore, we notice that $\min \left[(t^{k+1}_{11})^2, (t^{k+1}_{22})^2, (t^{k+1}_{33})^2 \right] \neq 0$ if $\min \left[(t^{k}_{11})^2, (t^{k}_{22})^2, (t^{k}_{33})^2 \right] \neq 0$ for all possible values of $\lambda^{k}$ $\in$ $(0, 1]$ and $\eta^{k}$ $\in$ $(0, 1]$.

When $\lambda^k = \eta^k = 1$, we have
\begin{align}
 t^{k+1}_{11} &= \frac{(t^k_{11})^{3}}{3 \left[(t^k_{11})^2 + (t^k_{22})^2 + (t^k_{33})^2\right]} ,  \nonumber \\
 t^{k+1}_{22} &= \frac{(t^k_{22})^3}{3 \left[(t^k_{11})^2 + (t^k_{22})^2 + (t^k_{33})^2\right]},   \nonumber \\
 t^{k+1}_{33} &= \frac{(t^k_{33})^3}{3 \left[(t^k_{11})^2 + (t^k_{22})^2 + (t^k_{33})^2\right]}.
\end{align}
Therefore, the state $\rho^{k+1}_{AB}$ remains to be Bell-diagonal with $\min \left[(t^{k+1}_{11})^2, (t^{k+1}_{22})^2, (t^{k+1}_{33})^2 \right] \neq 0$ if $\min \left[(t^{k}_{11})^2, (t^{k}_{22})^2, (t^{k}_{33})^2 \right] \neq 0$ even when Alice$^k$-Bob$^k$ performs projective measurements.

		\begin{table}[t]

\begin{tabular}{|c|c|c|c||c|c|c|c|}
\hline
\multicolumn{8}{|c|}{For $\rho_{AB}^1 = |\psi^{-}\rangle \langle \psi^{-} |$} \\
\hline
 Alice$^k$-Bob$^k$ & $\lambda^k$ & $\eta^k$ & $P_{\texttt{Min}}^{2 \rightarrow 1}$ &  Alice$^k$-Bob$^k$ & $\lambda^k$ & $\eta^k$ & $P_{\texttt{Min}}^{2 \rightarrow 1}$ \\
  with $k =$ &  &  &  & with $k =$  &  &  &     \\
 \hline 
 \hline
 $1$ & $1$ & $1$ & $0.854$ &  $1$ & $0.340$ & $0.340$ & $0.541$  \\
\hline
 $2$ & $1$ & $1$ & $0.588$ &  $2$ & $1$ & $1$ & $0.833$  \\
 \hline
 $3$ & $1$ & $1$ & $0.522$ &  $3$ & $1$ & $1$ & $0.583$  \\
 \hline
 $4, 5, 6, \cdots$  & $1$ & $1$ & $\frac{1}{2} < P_{\texttt{Min}}^{2 \rightarrow 1} < 0.520$ &  $4$ & $1$ & $1$ & $0.521$  \\
 \hline
   \multicolumn{4}{|c||}{} &   $5, 6, 7, \cdots$ & $1$ & $1$ & $\frac{1}{2} < P_{\texttt{Min}}^{2 \rightarrow 1} < 0.520$  \\
 \hline
\multicolumn{8}{c}{} \\
\multicolumn{8}{c}{} \\
 \hline 
\multicolumn{8}{|c|}{For $\rho_{AB}^1 = |\psi^{-}\rangle \langle \psi^{-} |$} \\
\hline
 Alice$^k$-Bob$^k$ & $\lambda^k$ & $\eta^k$ & $P_{\texttt{Min}}^{2 \rightarrow 1}$ &  Alice$^k$-Bob$^k$ & $\lambda^k$ & $\eta^k$ & $P_{\texttt{Min}}^{2 \rightarrow 1}$ \\
  with $k =$ &  &  &  & with $k =$  &  &  &     \\
 \hline 
 \hline
 $1$ & $0.340$ & $0.340$ & $0.541$ &  $1$ & $0.340$ & $0.340$ & $0.541$  \\
\hline
 $2$ & $0.340$ & $0.340$ & $0.538$ &  $2$ & $0.340$ & $0.340$ & $0.538$  \\
 \hline
 $3$ & $1$ & $1$ & $0.813$ &  $3$ & $0.340$ & $0.340$ & $0.536$  \\
 \hline
 $4$  & $1$ & $1$ & $0.578$ &  $4$ & $0.340$ & $0.340$ & $0.534$  \\
 \hline
   $5$  & $1$ & $1$ & $0.520$  &   $5$ & $0.340$ & $0.340$ & $0.532$  \\
   \hline
$6, 7, 8, \cdots$ & $1$ & $1$ &  $\frac{1}{2} < P_{\texttt{Min}}^{2 \rightarrow 1} < 0.520$ & $6$  & $0.340$ & $0.340$ & $0.530$  \\
 \hline
   \multicolumn{4}{|c||}{} &    $7$ & $0.340$ & $0.340$ & $0.528$  \\
   \cline{5-8}
   \multicolumn{4}{|c||}{} &   $8$ & $0.340$ & $0.340$ & $0.527$  \\
   \cline{5-8}
  \multicolumn{4}{|c||}{} &    $9$ & $0.340$ & $0.340$ & $0.525$  \\
   \cline{5-8}
   \multicolumn{4}{|c||}{} &    $10$ & $0.340$ & $0.340$ & $0.524$  \\
   \cline{5-8}
   \multicolumn{4}{|c||}{} &    $11$ & $0.340$ & $0.340$ & $0.522$  \\
   \cline{5-8}
   \multicolumn{4}{|c||}{} &   $12$ & $0.340$ & $0.340$ & $0.521$  \\
   \cline{5-8}
   \multicolumn{4}{|c||}{} &  $13$ & $0.340$ & $0.340$ & $0.520$  \\
   \cline{5-8}
  \multicolumn{4}{|c||}{} &   $14, 15, 16, \cdots$ & $0.340$ & $0.340$ & $\frac{1}{2} < P_{\texttt{Min}}^{2 \rightarrow 1} < 0.520$  \\
   \hline
\end{tabular}
\caption{Minimum success probabilities in $2 \rightarrow 1$ RAC task (assisted with two bits, shared from a common source and having maximally mixed marginal  at the receiver's end) by different consecutive pairs for different choices of $\lambda^k$, $\eta^k$ ($k \in \{1, 2, \cdots\}$). In each of the above tables, the initially shared two-qubit state is assumed to be singlet state, $\rho_{AB}^1 = |\psi^{-}\rangle \langle \psi^{-} |$. Here, all the numerical values are rounded to the third decimal places.} \label{tabu1}
\end{table}

	\section{Trade-off between the amount of quantum advantage gained by an arbitrary pair and the number of subsequent pairs exhibiting considerable amount of quantum advantage}\label{appendix4}

	We have shown that an unbounded pairs of Alice and Bob can, in principle, demonstrate quantum advantage in some particular $n \rightarrow 1$ RAC task (with $n=2$ or $n=3$). Here, quantum advantage implies that the magnitude of $(P_{\texttt{Min}}^{n \rightarrow 1} - \frac{1}{2})$ is positive. However, for experimental implementation, the magnitude of $(P_{\texttt{Min}}^{n \rightarrow 1} - \frac{1}{2})$ should not only be positive, but also be large enough to be detected in real experiment. In order to explore this issue relevant for practical realization, we consider that the quantum advantage is `significant' if $P_{\texttt{Min}}^{n \rightarrow 1} \geq 0.520$. Note, here the bound $0.520$ is chosen as an example. One can choose any other bound depending on the precision of the experimental apparatus and the characteristics of the results described below will not change. 
	
	We observe that when an arbitrary pair gains a large amount of quantum advantage by performing projective measurements or unsharp measurements with the sharpness parameters being close to unity, then only few number of subsequent pairs will get `significant' quantum advantage. On the other hand, when a pair gets a small amount of quantum advantage by performing unsharp measurements with the sharpness parameters being much less than unity, a larger number of subsequent pairs can achieve `significant' quantum advantage. We will describe this aspect through a number of examples. 
	
	Let us consider that the singlet state is initially shared by the pair Alice$^1$-Bob$^1$ and each of the multiple pairs performs $2 \rightarrow 1$ RAC task assisted with two bits, shared from a common source and having maximally mixed marginal  at the receiver's end. The results in this case are presented in Table \ref{tabu1}. From this table, we observe that at most three consecutive pairs can show $P_{\texttt{Min}}^{2 \rightarrow 1} \geq 0.520$ when all observers perform projective measurements. Next, consider the case when the first pair performs measurements with $\lambda^1 = 0.340$ and $\eta^1=0.340$ and all other pairs perform projective measurements. In this case, the minimum success probability for the first pair is reduced compared to the previous case (with $\lambda^1 = 1$ and $\eta^1=1$), but the maximum number of pairs exhibiting $P_{\texttt{Min}}^{2 \rightarrow 1} \geq 0.520$ is increased to four. Now, consider that the first and second pairs perform unsharp measurements with $\lambda^1 = 0.340$, $\eta^1=0.340$, $\lambda^2 = 0.340$, $\eta^2=0.340$ and all other pairs perform projective measurements. In this case, the minimum success probability for the second pair is  reduced compared to the case with $\lambda^2 = 1$ and $\eta^2=1$, and the maximum number of pairs exhibiting $P_{\texttt{Min}}^{2 \rightarrow 1} \geq 0.520$ is further increased to five. Proceeding in this way, we find out that when all the pairs perform measurements with $\lambda^k = 0.340$ and $\eta^k=0.340$ ($k \in \{1, 2, \cdots\}$), then the maximum number of pairs exhibiting $P_{\texttt{Min}}^{2 \rightarrow 1} \geq 0.520$ is increased to thirteen. 
	
	
	Next, let us focus on the $3 \rightarrow 1$ RAC task assisted with two bits shared from a common source. Here also, consider that the singlet state is initially shared by the pair Alice$^1$-Bob$^1$. The results in this case are presented in Table \ref{tabu2}. From this table, it can be noticed that at most two consecutive pairs can achieve $P_{\texttt{Min}}^{3 \rightarrow 1} \geq 0.520$ contingent upon performing projective measurements by all pairs. Next, consider that the first pair performs unsharp measurements associated with $\lambda^1 = 0.370$ and $\eta^1=0.370$ and each of the other pairs performs projective measurements. In this case, we observe that the minimum success probability for the first pair is reduced compared to the previous case with $\lambda^1 = 1$ and $\eta^1=1$, but the maximum number of pairs exhibiting $P_{\texttt{Min}}^{3 \rightarrow 1} \geq 0.520$ is increased to three. Next, we consider that the first pair and second pair perform unsharp measurements with $\lambda^1 = 0.370$, $\eta^1=0.370$ and $\lambda^2 = 0.370$, $\eta^2=0.370$ respectively. All other pairs perform projective measurements. In this case, the minimum success probability for the second pair is  decreased compared to the case with $\lambda^2 = 1$ and $\eta^2=1$. On the other hand, the maximum number of pairs exhibiting $P_{\texttt{Min}}^{3 \rightarrow 1} \geq 0.520$ is further increased to four. Following this approach, we find out that when all the pairs perform unsharp measurements with $\lambda^k = 0.370$ and $\eta^k=0.370$ for all $k \in \{1, 2, \cdots\}$,  the maximum number of pairs exhibiting $P_{\texttt{Min}}^{3 \rightarrow 1} \geq 0.520$ becomes eight. 
	
	It can be easily checked that the above aspects remain unaltered when the sequential execution of the aforementioned $n \rightarrow 1$ RAC task (with $n=2$ or $n=3$) is initiated with an appropriate Bell-diagonal two-qubit (entangled or separable) state given by Eq.(\ref{belldiam}) with $k=1$ and $\min\limits_{i \leq n} \left[ (t^1_{ii})^2 \right] \neq 0$ and for different choices of the sharpness parameters. Therefore, we can conclude that there may exist a trade-off relation between the amount of quantum advantage gained by an arbitrary pair and the number of subsequent pairs exhibiting $P_{\texttt{Min}}^{n \rightarrow 1} \geq 0.520$. Importantly, any of these two quantities can be increased at the expense of the other by suitably choosing the sharpness parameters associated with the measurements.

		\begin{table}[t]

\begin{tabular}{|c|c|c|c||c|c|c|c|}
\hline
\multicolumn{8}{|c|}{For $\rho_{AB}^1 = |\psi^{-}\rangle \langle \psi^{-} |$} \\
\hline
 Alice$^k$-Bob$^k$ & $\lambda^k$ & $\eta^k$ & $P_{\texttt{Min}}^{3 \rightarrow 1}$ &  Alice$^k$-Bob$^k$ & $\lambda^k$ & $\eta^k$ & $P_{\texttt{Min}}^{3 \rightarrow 1}$ \\
  with $k =$ &  &  &  & with $k =$  &  &  &     \\
 \hline 
 \hline
 $1$ & $1$ & $1$ & $0.789$ &  $1$ & $0.370$ & $0.370$ & $0.540$  \\
\hline
 $2$ & $1$ & $1$ & $0.532$ &  $2$ & $1$ & $1$ & $0.762$  \\
 \hline
 $3, 4, 5, \cdots$ & $1$ & $1$ & $\frac{1}{2} < P_{\texttt{Min}}^{3 \rightarrow 1} < 0.520$ &  $3$ & $1$ & $1$ & $0.529$  \\
 \hline
 \multicolumn{4}{|c||}{}  &  $4, 5, 6, \cdots$ & $1$ & $1$ & $\frac{1}{2} < P_{\texttt{Min}}^{3 \rightarrow 1} < 0.520$  \\
 \hline
\multicolumn{8}{c}{} \\
\multicolumn{8}{c}{} \\
 \hline 
\multicolumn{8}{|c|}{For $\rho_{AB}^1 = |\psi^{-}\rangle \langle \psi^{-} |$} \\
\hline
 Alice$^k$-Bob$^k$ & $\lambda^k$ & $\eta^k$ & $P_{\texttt{Min}}^{3 \rightarrow 1}$ &  Alice$^k$-Bob$^k$ & $\lambda^k$ & $\eta^k$ & $P_{\texttt{Min}}^{3 \rightarrow 1}$ \\
  with $k =$ &  &  &  & with $k =$  &  &  &     \\
 \hline 
 \hline
 $1$ & $0.370$ & $0.370$ & $0.540$ &  $1$ & $0.370$ & $0.370$ & $0.540$  \\
\hline
 $2$ & $0.370$ & $0.370$ & $0.536$ &  $2$ & $0.370$ & $0.370$ & $0.536$  \\
 \hline
 $3$ & $1$ & $1$ & $0.738$ &  $3$ & $0.370$ & $0.370$ & $0.532$  \\
 \hline
 $4$  & $1$ & $1$ & $0.526$ &  $4$ & $0.370$ & $0.370$ & $0.530$  \\
 \hline
   $5, 6, 7,  \cdots$  & $1$ & $1$ & $\frac{1}{2} < P_{\texttt{Min}}^{3 \rightarrow 1} < 0.520$  &   $5$ & $0.370$ & $0.370$ & $0.527$  \\
   \hline
\multicolumn{4}{|c||}{} & $6$  & $0.370$ & $0.370$ & $0.524$  \\
 \cline{5-8}
   \multicolumn{4}{|c||}{} &    $7$ & $0.370$ & $0.370$ & $0.522$  \\
   \cline{5-8}
   \multicolumn{4}{|c||}{} &   $8$ & $0.370$ & $0.370$ & $0.520$  \\
   \cline{5-8}
  \multicolumn{4}{|c||}{} &    $9, 10, 11,  \cdots$ & $0.370$ & $0.370$ & $\frac{1}{2} < P_{\texttt{Min}}^{3 \rightarrow 1} < 0.520$  \\
   \hline
\end{tabular}
\caption{Minimum success probabilities in $3 \rightarrow 1$ RAC task (assisted with two bits shared from a common source) by different consecutive pairs for different choices of $\lambda^k$, $\eta^k$ ($k \in \{1, 2, \cdots\}$). In each of the above tables, the initially shared two-qubit state is assumed to be singlet state, $\rho_{AB}^1 = |\psi^{-}\rangle \langle \psi^{-} |$. Here, all the numerical values are rounded to the third decimal places.} \label{tabu2}
\end{table}
	

	\section{Proof of Corollary \ref{mintheo3} and Corollary \ref{cor2}}\label{appendix5}
Suppose Alice$^k$-Bob$^k$ performs the aforementioned $2 \rightarrow 1$ RAC task with probability $p_k$ and the aforementioned $3 \rightarrow 1$ RAC with probability $1-p_k$ (with $0 \leq p_k \leq 1$) using the following Bell-diagonal two-qubit state,
	\begin{equation}
	 \rho^k_{AB}=\frac{1}{4}(\mathbf{I}_4 +\sum_{i=1}^{3}t^k_{ii} \, \sigma_{i} \otimes  \sigma_{i}),   
	 \label{belldia2}
	\end{equation} 
	where  $(t^k_{uu})^2 \geq (t^k_{vv})^2 \geq (t^k_{ww})^2 > 0$ for an arbitrary choice of $u \neq v \neq w \in \{1, 2, 3\}$.	This implies that $\min \left[(t^{k}_{11})^2, (t^{k}_{22})^2 \right] \neq 0$ and $\min \left[(t^{k}_{11})^2, (t^{k}_{22})^2, (t^{k}_{33})^2 \right] \neq 0$. Hence, while performing the $2 \rightarrow 1$ RAC, this state will give $P_{\texttt{Min}}^{2\rightarrow 1} > \frac{1}{2}$. On the other hand, while performing the $3 \rightarrow 1$ RAC, this state will give $P_{\texttt{Min}}^{3\rightarrow 1} > \frac{1}{2}$. Hence, overall, Alice$^k$-Bob$^k$ will achieve quantum advantage with this state.
	
	Next, the average post-measurement state $\rho_{AB}^{k+1}$ received by Alice$^{k+1}$-Bob$^{k+1}$ from Alice$^k$-Bob$^k$ is given by,
	\begin{equation}
\rho^{k+1}_{AB} = p_k \, \rho^{k+1}_{AB_{2 \rightarrow 1}} 	+   (1-p_k) \, \rho^{k+1}_{AB_{3 \rightarrow 1}},  
\label{convexsm}
	\end{equation}
where the average is taken since each pair act independently of other pairs. In the above equation, 	$\rho^{k+1}_{AB_{2 \rightarrow 1}}$ is the average post-measurement state received by Alice$^{k+1}$-Bob$^{k+1}$ when Alice$^k$-Bob$^k$ performs the aforementioned $2 \rightarrow 1$ RAC with certainty. One can evaluate $\rho^{k+1}_{AB_{2 \rightarrow 1}}$ using Eq.(\ref{postappbsm}). On the other hand, $\rho^{k+1}_{AB_{3 \rightarrow 1}}$ is the average post-measurement state received by Alice$^{k+1}$-Bob$^{k+1}$ when Alice$^k$-Bob$^k$ performs the aforementioned $3 \rightarrow 1$ RAC with certainty. It can be evaluated using Eq.(\ref{postappb2}). Hence, we can infer that the state $\rho_{AB}^{k+1}$ has the following Bell-diagonal form,
\begin{equation}
	 \rho^{k+1}_{AB}=\frac{1}{4}(\mathbf{I}_4 +\sum_{i=1}^{3} t^{k+1}_{ii} \, \sigma_{i} \otimes  \sigma_{i}),   
	 \label{pmssme}
	\end{equation} 
where $t^{k+1}_{ii}$	for $i=1,2,3$ can be calculated using Eqs.(\ref{tevsm2}), (\ref{tevsm4}), (\ref{convexsm}). Hence, we have 
\begin{align}
 t^{k+1}_{11} =& p_k \frac{t^k_{11} \, \left[1+ \sqrt{1-(\eta^k)^2}\right] \left[(t^k_{11})^2 + (t^k_{22})^2 \sqrt{1- (\lambda^k)^2} \right]}{2 \left[(t^k_{11})^2 + (t^k_{22})^2\right]} \nonumber \\
 &\quad + (1-p_k) \frac{t^k_{11} \, \left[1+ 2 \sqrt{1-(\eta^k)^2}\right] \left[(t^k_{11})^2 + \left[ (t^k_{22})^2 + (t^k_{33})^2 \right] \sqrt{1- (\lambda^k)^2} \right]}{3 \left[(t^k_{11})^2 + (t^k_{22})^2 + t^k_{33})^2\right]},  \\
 t^{k+1}_{22} =& p_k \frac{t^k_{22} \, \left[1+ \sqrt{1-(\eta^k)^2} \right]  \left[(t^k_{22})^2 + (t^k_{11})^2 \sqrt{1-(\lambda^k)^2}\right]}{2 \left[(t^k_{11})^2 + (t^k_{22})^2\right]}   \nonumber \\
 & \quad + (1-p_k) \frac{t^k_{22} \, \left[1+ 2 \sqrt{1-(\eta^k)^2}\right] \left[(t^k_{22})^2 + \left[ (t^k_{33})^2 + (t^k_{11})^2 \right] \sqrt{1- (\lambda^k)^2} \right]}{3 \left[(t^k_{11})^2 + (t^k_{22})^2 + t^k_{33})^2\right]}, \\
 t^{k+1}_{33} =& p_k t^k_{33} \, \sqrt{1-(\eta^k)^2} \sqrt{1-(\lambda^k)^2} + (1-p_k) \frac{t^k_{33} \, \left[1+ 2 \sqrt{1-(\eta^k)^2}\right] \left[(t^k_{33})^2 + \left[ (t^k_{11})^2 + (t^k_{22})^2 \right] \sqrt{1- (\lambda^k)^2} \right]}{3 \left[(t^k_{11})^2 + (t^k_{22})^2 + t^k_{33})^2\right]}.
 \label{tevsm5}
\end{align}
From the above equations, it is observed that when $p_k \in [0, 1)$, then $t_{11}^{k+1} \neq 0$, $t_{22}^{k+1} \neq 0$, $t_{33}^{k+1} \neq 0$ for all possible values of $\lambda^{k}$ $\in$ $(0, 1]$ and $\eta^{k}$ $\in$ $(0, 1]$. Hence, in this case, Alice$^{k+1}$-Bob$^{k+1}$ will get quantum advantage in both $2 \rightarrow 1$ RAC and $3 \rightarrow 1$ RAC. Now, suppose that Alice$^{k+1}$-Bob$^{k+1}$ performs $2 \rightarrow 1$ RAC task with probability $p_{k+1}$ and $3 \rightarrow 1$ RAC with probability $1-p_{k+1}$ (with $0 \leq p_{k+1} \leq 1$). Following the aforementioned argument, we can, therefore, conclude that Alice$^{k+1}$-Bob$^{k+1}$ will also get quantum advantage overall.

Next, consider that $p_k = 1$. In this case, $t_{11}^{k+1} \neq 0$, $t_{22}^{k+1} \neq 0$, for all possible values of $\lambda^{k}$ $\in$ $(0, 1]$ and $\eta^{k}$ $\in$ $(0, 1]$. But $t_{33}^{k+1} \neq 0$ only for $\lambda^{k}$ $\in$ $(0, 1)$ and $\eta^{k}$ $\in$ $(0, 1)$; $t_{33}^{k+1} = 0$ if $\lambda^{k} = 1$ and $\eta^{k} = 1$. Subsequently, Alice$^{k+1}$-Bob$^{k+1}$ will not get quantum advantage while performing $3 \rightarrow 1$ RAC if Alice$^{k}$-Bob$^{k}$ performs the $2 \rightarrow 1$ RAC with certainty (i.e., $p_k=1$) under projective measurements. Hence, in order to ensure overall quantum advantage by Alice$^{k+1}$-Bob$^{k+1}$ for all values of $p_{k+1} \in [0, 1]$, we need the following: $\lambda^{k} \neq 1$ and $\eta^{k} \neq  1$.  

To summarize, if both the pairs- Alice$^{k}$-Bob$^{k}$ and Alice$^{k+1}$-Bob$^{k+1}$ perform  $2 \rightarrow 1$ RAC task with some non-zero probability and $3 \rightarrow 1$ RAC with some non-zero probability, then both the pairs  will get overall quantum advantage for all values of  $\lambda^{k}$ $\in$ $(0, 1]$ and $\eta^{k}$ $\in$ $(0, 1]$.   Applying this argument to arbitrary number of pairs with $p_k$ being equal to or not equal to $p_{\tilde{k}}$ for all choices of $k \neq \tilde{k} \in \{1, 2, \cdots\}$, we get Corollary \ref{mintheo3}.

On the other hand, if Alice$^{k}$-Bob$^{k}$ performs  $2 \rightarrow 1$ RAC task with certainty, then we have $p_k = 1$. In this case, if  Alice$^{k+1}$-Bob$^{k+1}$ performs $3 \rightarrow 1$ RAC with certainty, then this pair will  not get quantum advantage if $\lambda^{k} = 1$ and $\eta^{k} = 1$. However, for all values of  $\lambda^{k}$ $\in$ $(0, 1)$ and $\eta^{k}$ $\in$ $(0, 1)$, Alice$^{k+1}$-Bob$^{k+1}$ will get quantum advantage. Applying this argument to arbitrary number of pairs, we get Corollary \ref{cor2}.

\end{widetext}
\end{document}